\documentclass[twocolumn,showpacs,preprintnumbers,amsmath,amssymb,prb]{revtex4}
\usepackage{graphicx} % Include figure files
\usepackage{dcolumn} % Align table columns on decimal point
\usepackage{bm} % bold math

\begin{document}

\title{\bf Influence of dephasing process on the quantum Hall effect
and the spin Hall effect}

\author{Yanxia Xing$^1$, Qing-feng Sun$^{1,\ast}$, and Jian Wang$^2$}

\address{$^1$Beijing National Laboratory for Condensed Matter Physics and
Institute of Physics, Chinese Academy of Sciences, Beijing 100080,
China\\
$^2$Department of Physics and the center of theoretical and
computational physics, The University of Hong Kong, Hong Kong, China
}

\begin{abstract}

We study the influence of the phase relaxation process on Hall
resistance and spin Hall current of a mesoscopic two-dimensional
(2D) four-terminal Hall cross-bar with or without Rashba spin-orbit
interaction (SOI) in a perpendicular uniform magnetic field. We find
that the plateaus of the Hall resistance with even number of edge
states can survive for very strong phase relaxation when the system
size becomes much longer than the phase coherence length. On the
other hand, the odd integer Hall resistance plateaus arising from
the SOI are easily destroyed by the weak phase relaxation during the
competition between the magnetic field and the SOI which delocalize
the edge states. In addition, we have also studied the transverse
spin Hall current and found that it exhibits resonant behavior
whenever the Fermi level crosses the Landau band of the system. The
phase relaxation process weakens the resonant spin Hall current and
enhances the non-resonant spin Hall current.

\end{abstract}

\maketitle

\section{introduction}

When the 2D electron system is subjected to a strong perpendicular
magnetic field, the energy spectrum becomes a series of impurity
broadened Landau bands with extended state in the center
\cite{Khmelnitskii,Laughlin} of each Landau band and the localized
state at the band edges. This gives rise to the integer quantum Hall
effect (IQHE),\cite{book,QHE} in which the Hall conductance is
quantized and jumps from one quantized value to another when the
Fermi energy sweeps through the impurity broadened bands or the
extended-states. Experimentally,\cite{experiment} the quantized unit
of Hall conductance $h/2e^2$ can even be specified in parts per
million, which becomes a resistance standard that is insensitive to
the particular sample and details of its fabrication, because of the
spatial separation in the transport states with the opposite
velocity and consequently the incredibly long mean free path in the
quantum Hall sample. Since the IQHE is discovered in 1980, it has
been extensively studied in the past several
decades,\cite{He,Mineev,FQH,Laughlin2,Read} and the many
characteristics of IQHE has been well understood now.

In addition to the quantum Hall effect, the spin Hall effect (in
which the longitudinal electronic field induces the transverse spin
current), especially the {\it intrinsic} spin Hall effect (SHE) has
recently been intensively studied. Different from the extrinsic SHE
which is due to the spin dependent scattering,\cite{Hirsch} the
intrinsic SHE is originated from spin-orbit interaction (SOI). It
was predicted first by Murakami {\it et.al.}\cite{Zhangsc} and
Sinova {\sl et.al.}\cite{Sinova} in a Luttinger spin-orbit coupled
3D p-doped semiconductor and a Rashba spin-orbit coupled 2D electron
gas, respectively. Subsequently, many related works are focused on
intrinsic
SHE.\cite{experiment1,experiment2,aref1,Sheng1,Shen,Xing1,Nikolic,Xing2,Qiao}
Sheng {\it et al.}\cite{Sheng1} investigated the SHE in the
mesoscopic 2D junction with Rashba SOI, and found that the SHE can
survive in the mesoscopic systems with weak disorder. Xing {\it et
al} \cite{Xing1} found that the intrinsic SHE is dominated by the
extend states, which is different from the IQHE. Besides, the
out-of-plane and in-plane component of transverse spin Hall current
was studied in a ballistic 2D finite electron
system.\cite{Nikolic,Xing2} On the experimental side, Kato {\sl et
al.}\cite{experiment1} and Wunderlich {\sl et al.}\cite{experiment2}
have observed the transverse opposite spin accumulations near two
edges of their devices when the longitudinal voltage bias is added.
In addition, the reversal SHE was observed by measuring an induced
transverse voltage in a diffusive metallic conductor when a
longitudinal net spin current flows through it.\cite{aref1}

Although the IQHE is insensitive to the particular sample and
details of its fabrication, it can be transformed to the insulating
regime with the weak magnetic field or strong impurity scattering.
In fact, global phase diagram of transitions between the quantum
Hall states and the insulator (or localized) state has been studied
for the quantum Hall system in the tight-binding model \cite{Xie}
and in 2DEG model.\cite{Kivelson} A phase diagram for the mesoscopic
SHE has also been proposed by Qiao {\it et al.}.\cite{Qiao} These
works showed both IQHE and SHE can survive in weak disorders and
IQHE is more robust than SHE in resisting impurity scatterers.
However, in a realistic sample, there exists both impurities or
rigid scatters that maintains the phase coherence and the dynamic
scatterers like lattice vibration (photons) and electron-electron
interactions that induce the phase-relaxation (PR). Hence it is
interesting to ask to what extent the IQHE or SHE can survive in the
presence of PR processes?

In this paper, we study influence of the PR processes on the IQHE
and SHE based on non-equilibrium Green's function (NEGF). We
consider a 2D mesoscopic device which is sketched in the inset of
Fig.1(c): The central square ballistic region is connected to the
four ideal semi-infinite lead-1,2,3 and 4 with the width $W$. The
whole system, including the central region and the four leads, lies
in the $x,y-$plane. A magnetic field $B_z$ is applied in the
positive $z-$direction. The PR processes in the central region are
phenomenologically simulated by introducing the virtual
leads\cite{virtual}, through which the electrons lose their phase
memory. This method was first introduced by B$\ddot{u}$ttiker in
1986. Moreover, there was another method to mimic the dephasing
process by Datta and his co-worker.\cite{virtual2} This method
provides a NEGF-based phenomenological model that is comparable to
the virtual leads method with conceptual and numerical simplicity.
In the following we will use the virtual leads method (also known as
B$\ddot{u}$ttiker probe model in some references) to mimic the
dephasing processes. For the system without SOI, the longitudinal
current or conductance $J_1$ (we have set the bias $V_1-V_3=1$) in
the lead-1, transverse Hall voltage $V_H=V_2-V_4$ and the step-like
Hall resistance $\rho_H=V_H/J_1$ are calculated numerically with the
aid of Green's function method. The results show that IQHE can
survive at strong PR process. In particular, the quantized plateaus
of the Hall conductance can be kept well even when the PR process is
so strong as to completely relax the transport current. In the
presence of the SOI, spin degeneracy is broken and odd integer Hall
plateaus emerge. These odd number edge states are easily destroyed
by weak PR processes. In addition, we also investigate how the SHE
is affected by the PR process. It is found that the spin Hall
currents $J_{2/4,s}$ show the resonant behaviors when the filling
factor changes from odd to even where the Fermi energy is in line
with a branch of the eigen levels of the spin polarized system.
Furthermore, PR processes weaken the resonant spin current but
enhance non-resonant spin current.

The rest of this paper is organized as follows: In Sec. II, the
system Hamiltonian and the theoretical formula for calculating the
Hall resistance and other quantities are presented. In Sec. III,
we show the numerical results and some discussions. Finally, a
brief summary is given in Sec. IV.

\section{Hamiltonian and formula}

Our system (in the absence of phase relaxation and disorders) can be
described by the Hamiltonian $H=(-i\hbar\nabla+e{\bf
A})^2/2m^*+\alpha(\sigma_x \nabla_y-\sigma_y \nabla_x)$ with $e$ the
electron charge, $m^*$ the effective mass, $\alpha$ the strength of
the Rashba SOI, and $\sigma_{x,y}$ the Pauli matrix. Generally
speaking, there are always impurities or disorders in the realistic
system giving rise to rigid scatterings which do not contribute to
phase-relaxation process. This type of impurity scattering has been
extensively investigated using the model of the on-site white-noise
potential $V_{\bf i}$ ({\bf i} denotes the lattice site) distributed
uniformly from -D/2 to D/2.\cite{Sheng1,Xing1,Qiao} On the other
hand, in the presence of phase-randomizing collisions, such as the
dynamic scattering by the lattice vibrations (phonon), the
electron-electron scattering, and so on, the transport electron
loses the phase memory due to these PR processes. Although a proper
treatment of non-coherent transport requires advanced delicate
concepts, the basic issues can be accounted for by introducing the
virtual leads\cite{virtual} attached to the site ${\bf i}$ to mimic
the phase-breaking process occurring at the site ${\bf
i}$.\cite{virtual} Then the tunneling electrons can escape from the
site ${\bf i}$ into the virtual leads where the electrons lose phase
memory completely and finally return back to the site ${\bf i}$. We
assume that the PR processes occur only in the central region and
the leads are treated as the measurement terminal which is ideal and
clean.

For the central and four real leads, we introduce the tight-binding
representation, and the virtual leads are assumed in the
free-electron form and expressed in the $k-$space, which is not
necessary but for simplicity. Then, the Hamiltonian is written in
the following form:\cite{Xing1,Qiao}
\begin{eqnarray}
H&=& -\sum\limits_{\bf i} \left[a_{\bf i}^\dagger (  t e^{-i m
\phi} \sigma_0-iV_R\sigma_y)a_{{\bf i}+\delta_x} \right.
\nonumber \\
&& + \left.a_{\bf i}^\dagger (t\sigma_0+iV_R\sigma_x) a_{{\bf i}+
\delta_y}+H.c \right] \nonumber \\
&+&\sum\limits_{\bf i} a_{\bf i}^\dagger [B_z g_s
\mu_B\sigma_z/2]a_{\bf i}\nonumber \\
&+&\sum\limits_{{\bf i},k} \left[ \epsilon_k a_{{\bf i}k}^\dagger
a_{{\bf i}k} + (t_k a_{\bf i}^\dagger  a_{{\bf i}k} + H.c) \right]
\end{eqnarray}
where the first term describes the nearest neighbor coupling and the
Rashba SOI in the central region and real leads, in which, ${\bf
i}=({\bf i}_x,{\bf i}_y)$ describes the site of the 2D region shown
in the inset of the Fig.1(c). $a_{{\bf i}}=[a_{{\bf
i},\uparrow},a_{{\bf i},\downarrow}]^T$ is the annihilation operator
of electrons on the lattice site {\bf i}, and $\delta x,\delta y$
are unit vectors along the x and y directions. $\sigma_{x,y,z}$ are
Pauli matrices and $\sigma_0$ is a $2\times 2$ unit matrix.
$t=\hbar^2/2m^*a^2$ is the nearest neighbor hopping matrix element
with the lattice constant $a$. $\phi=qB_z a^2/\hbar$ is the extra
phase unit originated from the vector potential ${\bf A}$ and m
comes from ${\bf i}_y=ma$. In the presence of the magnetic field
${\bf B}=(0,0,B_z)$, we introduce the vector potential ${\bf
A}=(-B_zy,0,0)$ in the Landau Gauge, so that the extra phase $\phi$
occurs in the form $e^{-im\phi}$ in the $x-$direction.
$V_R=\alpha/2a$ denotes the Rashba SOI strength in the tight-binding
representation. In our model, the magnetic field $B_z$ is uniform in
the whole system, including the four leads and the center region.
While the SOI strength $V_R$ exists only in the longitudinal
lead-1,3 and the central region, and $V_R$ is set to zero in the
transverse lead-2,4. The second term in Eq.(1) denotes the Zeeman
split where $B_z$, $g_s$ and $\mu_B$ are the magnetic field along
$z-$direction, the Land$\acute{e}$ $g$ factor, and Bohr magneton,
respectively. Finally, the last term in Eq.(1) represents the
Hamiltonian of the virtual leads (described in the $k-$space) and
their coupling to the central site ${\bf i}$. $a_{{\bf i}k}=[a_{{\bf
i}k,\uparrow},a_{{\bf i}k,\downarrow}]^T$ is the annihilation
operator of the electrons in the virtual leads, where ${\bf i}$
signs the positions of central region. Every central site is coupled
by a virtual lead, so there totally are $N=W \times W$ virtual
leads, with $N$ being the site number in the central region and $W$
being the width of the central region.

The charge current in real leads $J_{r}$ ($r=1$, $2$, $3$, $4$) and
in virtual leads $J_{v}$ ($v=1,2, ... N$) can be obtained from the
Landauer-B\"{u}ttiker formula: \cite{Dattabook}
\begin{equation}\label{Landau1}
J_{p}=\frac{e^2}{h}\sum_{q\neq p}T_{pq}(V_p-V_q),
\end{equation}
where $p,q \in r$ or $v$, $V_p$ is the bias in the lead-$p$ and
$T_{p,q}$ is the transmission coefficient from the lead-$q$ to the
lead-$p$. The transmission coefficient can be calculated from
$T_{pq}=Tr[{\bf \Gamma}_{p}{\bf G}^r{\bf \Gamma}_{q}{\bf G}^a]$,
where the line-width function ${\bf \Gamma}_{p}=i({\bf
\Sigma}_{p}^r-{\bf \Sigma}_{p}^{r\dagger})$ with ${\bf
\Sigma}_{p}^r$ the retarded self-energy and the Green's function
${\bf G}^r=[{\bf G}^a]^{\dagger}=\{E_F{\bf I}-{\bf H}_0-\sum_{p}{\bf
\Sigma}^r_{p}\}^{-1}$ where ${\bf I}$ is the unit matrix with the
same dimension as that of ${\bf H}_0$. In addition, in order to
investigate the SHE, we also need to calculate the spin Hall current
$J_{p,s}$ in the transverse lead-2 and 4 for the system with SOI.
Because we have set $V_R=0$ in lead-2 and 4 so that $\sigma$ is a
good quantum number, the particle current $J_{p\sigma}$ in the
lead-$p$ ($p=2$, $4$) with spin index $\sigma$ ($\sigma=\uparrow$ or
$\downarrow$) can be obtained from the Landauer-B\"{u}ttiker
formula:
\begin{equation}\label{Landau2}
J_{p\sigma}=\frac{e}{h}\sum_{q\neq p}T_{p\sigma,q}(V_p-V_q).
\end{equation}
The quantities in Eq.(\ref{Landau2}) are the same as that in
Eq.(\ref{Landau1}), except that here $T_{p\sigma,q}$ is the
transmission coefficient from the lead-$q$ to the lead-$p$ with spin
$\sigma$, and $T_{p\sigma,q}=Tr[{\bf \Gamma}_{p\sigma}{\bf G}^r{\bf
\Gamma}_{q}{\bf G}^a]$ with ${\bf \Gamma}_{p\sigma}=i({\bf
\Sigma}_{p\sigma}^r-{\bf \Sigma}_{p\sigma}^{r\dagger})$. After
obtaining the particle current $J_{p\sigma}$, the spin current
$J_{p,s}$ can be easily obtained as $J_{p,s}
=(\hbar/2)[J_{p\uparrow}-J_{p\downarrow}]$.

In our calculation, the external bias is applied in the longitudinal
leads-1 and 3 with $V_1=0.5$ and $V_3=-0.5$, thus electrons obtain a
velocity $v_x$ along the $x-$direction. With the perpendicular
magnetic field, the tunneling electrons deflect to the transverse
direction ($y$-direction) and generates the charge pile up at the
vicinity of the lead-2 and lead-4. It consequently leads to opposite
transverse Hall voltage $V_2=-V_4$ in the lead-2 and lead-4 which is
calculated by requiring the boundary condition $J_2=J_4=0$ since the
lead-2 and 4 only act as the voltage probes. Furthermore, the
electrons can only lose their phase memory by escaping into or
coming back from the virtual leads, and do not contribute net
current to the virtual leads, so there are $N$ extra boundary
conditions $J_v=0$ with $v=1,2,...N$. With the transmission
coefficient $T_{pq}$, the input parameters $V_1$ and $V_3$, and the
boundary condition $J_{p=2,4,v}=0$,  we can get the transverse bias
$V_{2,4}$ ($V_2=-V_4$) in the lead-2,4 and the longitudinal current
$J_{1,3}$ ($J_1=-J_3$) in the lead-1,3 using Eq.(\ref{Landau1}).
Consequently the Hall voltage $V_H=V_2-V_4$ and Hall resistance
$\rho_H=V_H/J_1$ are obtained straightforwardly. On the other hand,
when investigating the SHE we use the boundary condition $V_2=V_4=0$
instead of $J_2=J_4=0$. Then, the particle current $J_{p,\sigma}$
and consequently the spin current $J_{p,s}$ in the lead-2,4 can also
be calculated easily.

\section{numerical results and discussion}

In the numerical calculation, we fix the Fermi energy\cite{foot2}
$E_F=-3t$ which is near the band bottom $-4t$, and take $t$ as
energy unit. Here $t=\hbar^2/(2m^*a^2)$ is about $5meV$ while taking
$m^*=0.05m_e$ (the mass of free electron) and the lattice constant
$a=12.5nm$. The size of finite sample $W$ is confined to $W=40a$ and
$20a$. In the presence of magnetic field $B_z$, there is an extra
phase unit $\phi=\frac{q}{\hbar}Ba^2$ with the vector potential
${\bf A}=(-B_zy,0,0)$. When $\phi=1$, $B=\hbar/(ea^2)$, so we take
$\hbar/(ea^2)$ as the unit of the magnetic field $B$, which
corresponds to $B=4.2T$ and $\mu_B B= 0.05t$. We set Land$\acute{e}$
$g$ factor $g_s=2$. Thus Zeeman splitting $\frac{1}{2}g_s\mu_B B=
0.05t$. Moreover, in order to generate the edge states, the
cyclotron radius $r_c=v/\omega_c$ must satisfy $r_c<W/2$, then the
magnetic field $2/W<B$ is needed. In the SOI system, $V_R=xt$
corresponds to the strength of Rashba SOI
$\alpha=2aV_R\approx1.25x\times 10^{-10}eVm$ and spin precession
length (over which the precessing angle $\pi$ is generated)
$L_{SO}=\pi a/(2x)$. Finally, as a check for our computer program,
we have calculated the case in which the magnetic field and the PR
process is absent but the spin-orbit interaction is present, the
same result as in the Fig.1 of the first reference in Ref.14 can
been obtained.

\begin{figure}%[tbp]
\includegraphics[bb=8mm 9mm 205mm 171mm, width=8cm,totalheight=6.5cm,
clip=]{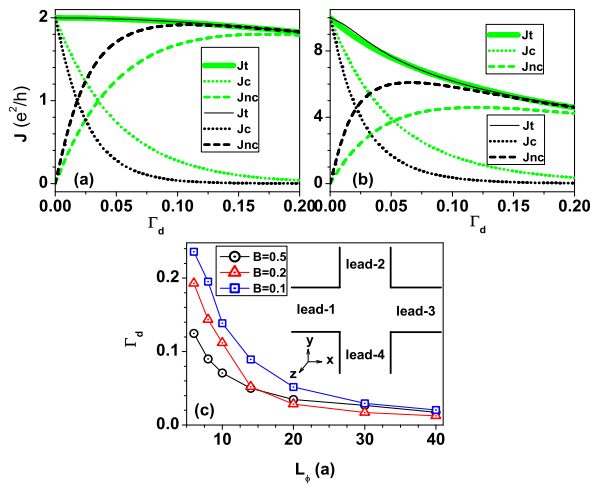} \caption{(Color online) The total current $J_t$,
the coherent component $J_c$, and the non-coherent component
$J_{nc}$ {\sl vs.} the PR strength $\Gamma_d$ in the two-terminal
non-SOI system for the magnetic field $B=0.5$ (panel (a)) and
$B=0.1$ (panel (b)) and the system size $W=20$ (green or gray lines)
and $W=40a$ (black lines). In the Panel (c), the PR strength
$\Gamma_d$ {\sl vs.} phase coherence length $L_\phi$ is plotted.
Inset of panel (c): Schematic diagram for the mesoscopic
four-terminal device.}
\end{figure}

In the experiment, the phase coherent length $L_\phi$ is a
observable parameter and is used to describe all kinds of dephasing
processes. Under different experimental conditions, there are
different dephasing processes and correspondingly different
$L_\phi$. So in the following, we will study the relation between
the PR strength $\Gamma_d$ and the phase coherent length $L_\phi$.
Here we define $L_\phi$ as the length, through which the transport
electron has 50\% probability to lose its phase memory and 50\%
probability to keep the phase coherence. In the presence of PR
processes $\Gamma_d$, the current generally consists of both phase
coherent part and phase incoherent part. In order to estimate the
phase coherence length $L_\phi$ for a given PR strength $\Gamma_d$,
we consider a two-terminal structure (decoupled with the lead-2,4),
in which the electrons can directly flow from lead-1 to lead-3 or
indirectly from lead-1 to lead-3 through virtual leads, the former
contributes to the phase coherent current and the latter the
incoherent part. Fig.1a and b show the coherent component, the
incoherent component, and the total current versus $\Gamma_d$ for
the different system sizes (i.e. the system lengths) $W=20a$ and
$W=40a$. The total current decreases slightly with the increasing
$\Gamma_d$, since the edge states carrying the transport electrons
are slightly destroyed to the extent proportional to $\Gamma_d$. At
$\Gamma_d=0$, the incoherent component is zero and the coherent
component is equal to the total current. With the increase of
$\Gamma_d$, the incoherent component increases and the coherent
component decreases. At a certain $\Gamma_d$, i.e., at the crossing
point of the two curves, the incoherent component is just equal to
the coherent component. This means that for this critical
$\Gamma_{dc}$ the system length $W$ is just equal to the phase
relaxation length $L_\phi$. Therefore the relation between $L_\phi$
and the PR strength $\Gamma_d$ can be obtained. Fig.1c shows
$\Gamma_d$ versus $L_\phi$ for different magnetic fields $B$. The
phase coherence length $L_\phi$ increases monotonically when
$\Gamma_d$ decreases. Obviously, when $\Gamma_d \rightarrow 0$,
$L_\phi \rightarrow \infty$. When $\Gamma_d$ is not very small (e.g.
$\Gamma_d>0.05t$), the bigger the magnetic field $B$, the longer
$L_\phi$ is. This is because for the stronger $B$ the edge states
are more robust against the PR process. In particular, for the given
system size $W=40a$ and $20a$ used in our following calculation, the
critical $\Gamma_{dc}\approx 0.02$ and $0.05$, respectively, beyond
which the phase coherence length is smaller than the system size.

\subsection{the IQHE}

In this subsection, we study the influence of the PR processes on
the IQHE. Firstly, we consider the system without Rashba SOI. In the
absence of PR processes ($\Gamma_d=0$), all of the electrons are
carried by the edge states, the electrons traverse clockwisely along
the sample edge due to the Lorentz force $qv\times B$ (here $B=B_z$
is along the positive $z$-direction), and the only nonzero matrix
elements of the transmission coefficient matrix $T$ are $T_{21},
T_{32}, T_{43}$ and $T_{14}$ with integer values. When PR processes
exist ($\Gamma_d\not=0$), the edge state is partially destroyed. As
a result, the elements $T_{21}, T_{32}, T_{43}$ and $T_{14}$ of the
transmission coefficients, which denote the edge states, deviate
from integer values, and the other elements such as $T_{12}$ become
nonzero.
\begin{figure}%[tbp]
\includegraphics[bb=9mm 10mm 203mm 190mm, width=8.5cm,totalheight=7cm,
clip=]{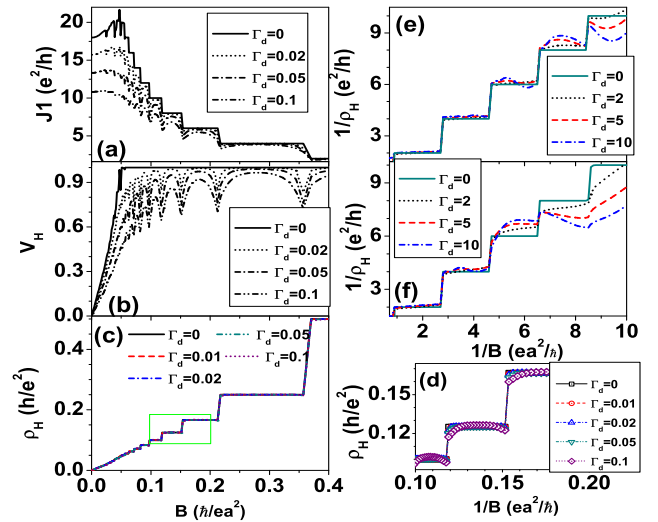} \caption{(Color online) Panel (a)-(c): the current
in the lead-1 $J_1$, the transverse Hall voltage $V_H$ and Hall
resistivity $\rho_H$ {\sl vs.} magnetic field $B$ for the different
PR strength $\Gamma_d$, with $V_R=0$ and $W=40a$. Panel (d)
magnifies the marked region in the panel (c). Inverse of Hall
resistivity $1/\rho_H$ {\sl vs.} the inverse of the magnetic field
$1/B$ in the non-SOI system ($V_R=0$) for the different system size
$W=40a$ (panel (e)) and $W=20a$ (panel (f)).}
\end{figure}
In Fig.2, the current $J_1$ in the lead-1, the transverse Hall
voltage $V_H$ and Hall resistivity $\rho_H$ or $1/\rho_H$ versus
magnetic field $B$ or $1/B$ are plotted for the different
$\Gamma_d$. From Fig.2, we can see when the magnetic field
($B<0.05$) is too small to form edge states, $J_1$ changes slowly
for the variable $B$ and decreases rapidly with $\Gamma_d$, the Hall
voltage $V_H$ and Hall resistance $\rho_H$ increase linearly with
$B$, which is in agreement with the results of semiclassical Drude
model. In the following, we focus on the high fields $B$ case, i.e.,
$B$ is large enough to separate the flows with the opposite
velocity,\cite{Datta} and the system is in the quantum Hall regime.
When $\Gamma_d=0$, the Hall voltage $V_H=V_2-V_4=1$, $J_1$ and
consequently $\rho_H$ exhibit plateaus. This can be understood using
the well known picture of Landau level (LL) and the edge state
\cite{Landau}. In the presence of PR process ($\Gamma_d\not=0$),
$J_1$ seriously deviates from the even integer plateaus and $V_H$ is
no longer a constant. However $\rho_H$ hardly changes and still
keeps plateaus even when $\Gamma_d$ is much bigger than the critical
value $\Gamma_{dc}$ (see Fig.2c and d). This means that although the
system is in strong PR regime, the IQHE can survive and is rather
robust against the PR processes. From Fig.2(d), we can see that the
the plateaus are first destroyed at the band edges for large
$\Gamma_d$. This is because at the band edges of plateaus, $E_F$ is
closer to LLs than at the band centers of plateaus. As a result, the
electrons are easier to be relaxed to LLs leading to a smaller
$\rho_H$. In Fig.2e and f, we plot $1/\rho_H$ versus $1/B$ for very
large $\Gamma_d$. The results show that it is more difficult to
destroy the plateau at larger magnetic field $B$ or larger sample
size. So those plateaus can survive at a bigger $\Gamma_d$, i.e.,
they have stronger ability to resist the PR processes. This is
because the energy spacing $\Delta E$ between the nearest LLs is
larger for the larger magnetic field $B$, and the edge states
carrying the opposite current are separated at a larger distance for
the larger sample. In particular, the first plateau at $1/B <3$ can
keep well even when $\Gamma_d$ reaches $10$ which is two orders of
magnitude larger than the critical value $\Gamma_{dc}=0.02$ or
$0.05$.

\begin{figure}%[tbp]
\includegraphics[bb=10mm 12mm 196mm 240mm, width=6cm,totalheight=7cm,
clip=]{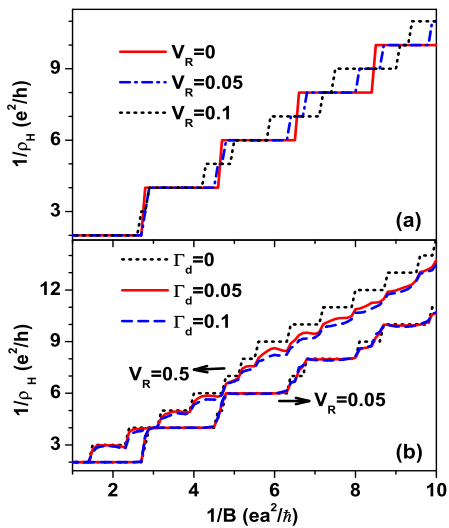} \caption{(Color online) Panel (a): inverse of Hall
resistivity $1/\rho_H$ {\sl vs.} the inverse of magnetic field $1/B$
for the different Rashba SOI strength $V_R$ at $\Gamma_d=0$. Panel
(b): $1/\rho_H$ {\sl vs.} $1/B$ with $V_R=0.05t$ and $0.5t$ for
different $\Gamma_d$. The system size $W=40a$.}
\end{figure}

Next, we study the system with Rashba SOI $V_R$. In Fig.3a,
$1/\rho_H$ versus $1/B$ for different $V_R$ in the absence of PR
processes ($\Gamma_d=0$) is plotted. It is interesting that although
the SOI induces the extended states, \cite{Xing1} $1/\rho_H$ is
still quantized and the integer quantum plateaus still remain. At
$V_R=0$, there are only even quanta for $1/\rho_H$. However, as long
as $V_R\neq0$, odd quanta emerge because the spin degeneracy is
destroyed. The width of the odd plateau is wider for the larger
Rashba SOI strength $V_R$ or for the smaller magnetic field $B$ (see
Fig.3a). While at the large $V_R$, the width of the odd plateau can
be in same order with, or even wider than, the width of the even
plateau.

In the following, let us study the influence of the PR processes on
the plateaus of the IQHE with non-zero Rahsba SOI ($V_R\not=0$).
Fig.3 shows $1/\rho_H$ versus $1/B$ for different SOI strengths
$V_R$ and PR strengths $\Gamma_d$. When $V_R$ is small, for example
$V_R=0.05t$, the even plateaus originated from the edge states are
still present. The odd plateaus due to SOI, however, are quickly
washed out. While for the strong SOI (e.g. $V_R=0.5t$) case, the SOI
that favors the extend states dominates the magnetic field that
favors the edge states, both even and odd plateaus are destroyed by
the weak PR process. Notice that for the non-SOI ($V_R=0$) case the
even plateaus can be kept at very large $\Gamma_d$ which can even
reach 10 as shown in Fig.2e and f. The fact that the even plateaus
in the SOI system are not as robust against the PR processes as that
in the non-SOI system, indicates that the SOI weakens the ability of
resisting the PR processes. In fact, the Rashba SOI coefficient
$\alpha$ is usually less than $10^{-11}eVm$ for a general 2D
electron gas, and the corresponding $V_R$ is less than $0.08t$. With
this $V_R$ (e.g. $V_R=0.05t$), the even plateaus can still survive
at $\Gamma_d =0.1$ which is much larger than the critical value
$\Gamma_{dc}$.

\begin{figure}%[tbp]
\includegraphics[bb=9mm 9mm 204mm 219mm, width=7cm,totalheight=7cm,
clip=]{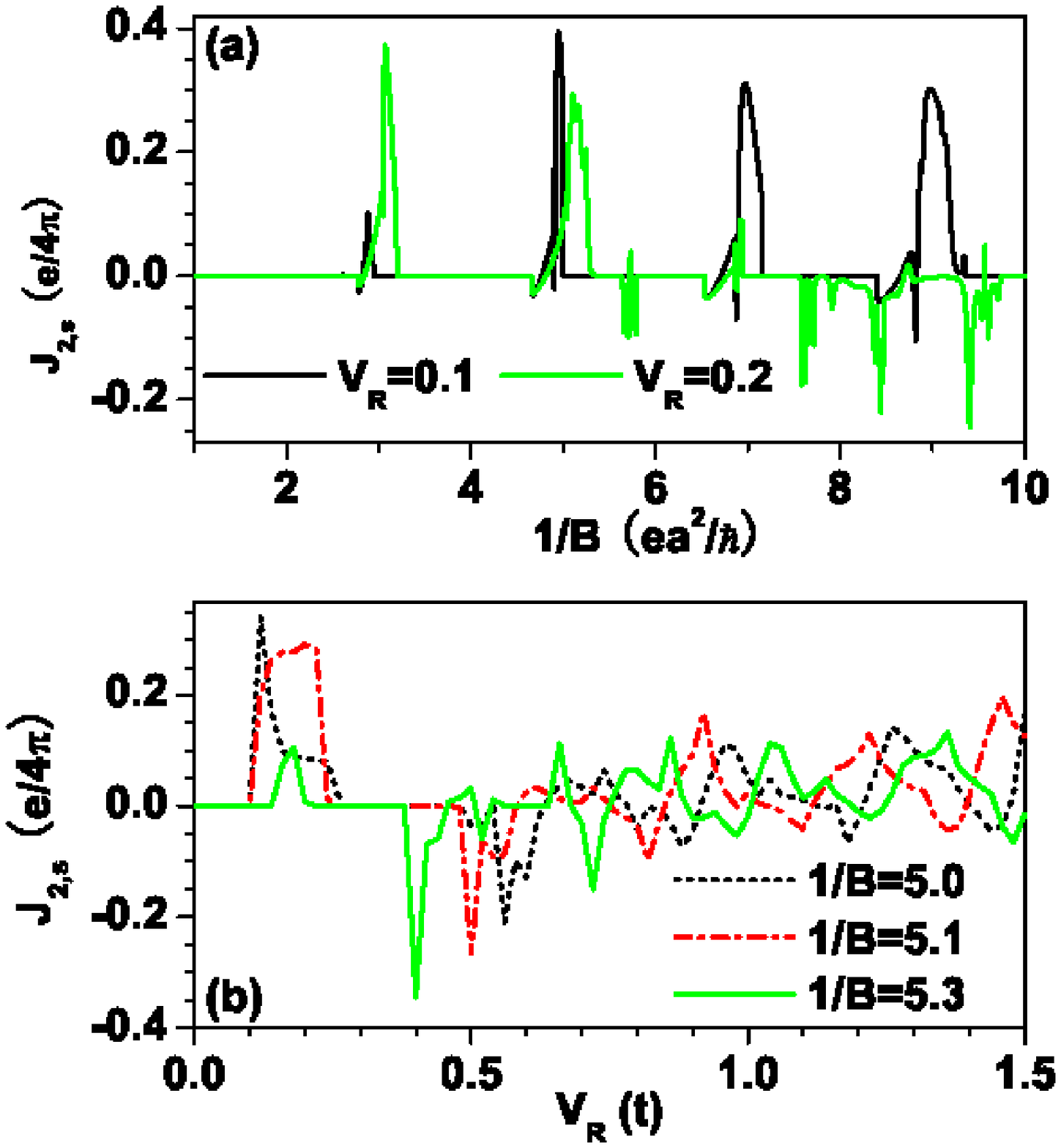} \caption{(Color online) Panel (a): The transverse
spin current $J_{2,s}$ {\sl vs.} inverse of magnetic field $1/B$ for
the different Rashba SOI strength $V_R$. Panel (b): The transverse
spin current $J_{2,s}$ {\sl vs.} the Rashba SOI strength $V_R$ near
the resonant peak $1/B\simeq 5.1$ in the panel (a). The other
parameters are $\Gamma_d=0$ and $W=40a$.}
\end{figure}

\subsection{the SHE}

In the system with SOI, the SHE occurs, in which a pure and
non-dissipating transverse spin current can be generated when a
longitudinal electric field or bias is applied. Recently, the SHE
has been extensively investigated by a great deal of works as
mentioned in the
introduction.\cite{experiment1,experiment2,aref1,Sheng1,Shen,Xing1,Nikolic,Xing2,Qiao}
Here we mainly study how the SHE is affected by the magnetic field
$B$, and, in particularly, the PR processes $\Gamma_d$. Firstly, the
case of $\Gamma_d=0$ is studied. The transverse spin current
$J_{2,s}=-J_{4,s}$ versus $1/B$ for different $V_R$ is plotted in
Fig.4(a). An interesting feature is that the spin current $J_{2,s}$
shows a resonant behavior, when the quantized $1/\rho_H$ changes
from the odd plateau to the even plateau where the Fermi level is in
line with the one of energy eigenvalues of the spin degenerated
system. The origin of the resonant spin current will be discussed at
last paragraph in this section (see Fig.6).
%
%It can be interpreted as follows: in the magnetic field subjected
%system without SOI, the eigen values are a spin degenerated Landau
%Level (LL) series. But in the presence of SOI, the LL series split
%into two branch levels. For the one branch levels, the
%corresponding states are strongly spin polarized. While for the
%other, there are hardly spin polarization. It is agree with the
%results of the Schr$\ddot{o}$dinger equation solving. For visual
%observation, we can turn to the Fig.6 in which we have signed the
%resonant positions through matching the Fermi energy and one of
%the LL series.
%
The spin Hall current $J_{2,s}$ is quite large when $1/B$ is near
the resonance, but is very small when $1/B$ is far away from it.
For $V_R=0.2$, the resonant positions are about $1/B=3.1t, 5.1t,
\cdots$, and for $V_R=0.1$, they are about at $1/B=3.0t,
5.0t,9.0t, 7.0t, \cdots$.
% In fact, when magnetic field
%is near the edge of plateau (see marks in Fig.4(a)), the current
%starts to polarize. The physics behind this can be understood as
%follows. Note that when the Fermi energy sweeps across bulk LLs
%the quantity $1/\rho_H$ jumps from one plateau to another in the
%absence of SOI. Also note that the effect of SOI is to polarize
%the spin of electrons and destroy the QHE. Therefore due to the
%SOI, it is easier for the electrons with energy closer to LLs to
%relax to the bulk LLs. As a result, the QHE is destroyed near the
%edge of plateau with large spin polarization (or large spin
%current) while is maintained near the center with no spin
%polarization (or zero spin current).
In Fig.4(b), we fix the
magnetic field $1/B$ at a resonant point ($1/B=5.1$) and plot
$J_{2,s}$ versus $V_R$. The results show that $J_{2,s}$ is
randomly distributed at the large $V_R$ because the extended
states are dominant. On the other hand, when $V_R<0.5t$, $J_{2,s}$
is regular. For $1/B=5.1$, $J_{2,s}$ is resonant for $V_R=0.2t$,
and there is a stable interval near $V_R=0.2t$ (see the red dash
dotted line in Fig.4b). When $1/B$ deviates from the resonant
point $5.1$, $J_{2,s}$ is rapidly decay from the resonant $V_R$
(see black dotted line and green solid line in Fig.4b).

\begin{figure}%[tbp]
\includegraphics[bb=9mm 9mm 210mm 141mm, width=8.5cm,totalheight=5.5cm,
clip=]{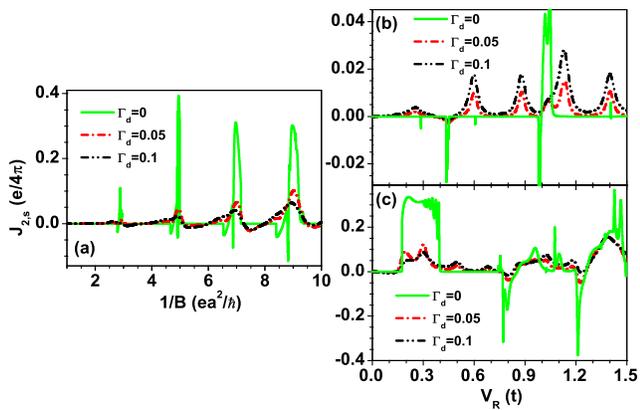} \caption{(Color online) Panel (a): the transverse
spin current $J_{2,s}$ {\sl vs.} inverse of magnetic field $1/B$ for
the different PR strength $\Gamma_d$ at $V_R=0.1$. Panel (b) and
(c): the transverse spin current $J_{2,s}$ {\sl vs.} the Rashba SOI
strength $V_R$ for the different PR strength $\Gamma_d$ at the
off-resonant peak $1/B=2$ (b) and resonant peak $1/B=3.1$ (c) in the
panel (a). The system size $W=40a$.}
\end{figure}

In the following, we study the influence of the PR processes
$\Gamma_d$ on the transverse spin current $J_{2,s}$. In Fig.5(a), we
plot $J_{2,s}$ vs the magnetic field $1/B$ for different $\Gamma_d$.
It shows that the PR process $\Gamma_d$ suppresses the spin current
$J_{2,s}$ when $1/B$ is near the resonant points, but enhances
$J_{2,s}$ in the off-resonant region. Next, picking up the
off-resonant position ($1/B=2$) and resonant position ($1/B=3.1$),
$J_{2,s}$ versus the SOI strength $V_R$ for different $\Gamma_d$ are
plotted in Fig.5b and c, respectively. Similar to Fig.5(a), the
off-resonant spin current is enhanced by the PR processes (see
Fig.5b). On the other hand, when the magnetic field is at the
resonant point (e.g. $1/B=3.1$ in Fig.5c), the resonant spin current
with $V_R$ from $0.2$ to $0.4$ is suppressed (see Fig.5c). Moreover,
when $V_R$ is very large (e.g. $V_R>0.5$), the spin current enters
the chaotic regime and it depends on the PR processes $\Gamma_d$ in
a random fashion.

\begin{figure}%[tbp]
\includegraphics[bb=13mm 11mm 197mm 227mm, width=6cm, totalheight=7cm,
clip=]{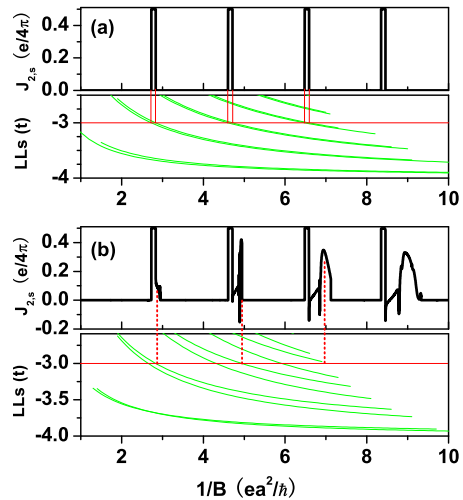} \caption{(color on line) The transverse spin
current $J_{2,s}$, the system Landau levels (LLs) {\sl vs.} inverse
of magnetic field $1/B$ while in the presence of the Zeeman effect,
with the Rashba SOI strength $V_R=0$ (a) and $V_R=0.1t$ (b). The
other parameters are $\Gamma_d=0$ and $W=40a$. The Fermi level is
fixed at $E=-3t$.}
\end{figure}

Finally, we also study the influence of the Zeeman effect on the
SHE. Fig.6(a) and (b) show the spin current $J_{2,s}$ versus the
inverse of magnetic field $1/B$ with the SOI strength $V_R=0$ and
$0.1t$, respectively. In each panel, we also plot the corresponding
eigenvalues (LLs) of the central scattering region (without leads)
versus $1/B$ and assign the resonant positions (see the red dotted
lines). When $V_R=0$, there is only one group of peaks in the curve
of $J_{2,s}$ vs $1/B$ due to the Zeeman effect. While for
$V_R\neq0$, the peaks appear in two groups corresponding to the
Zeeman peak and resonant peak of spin current. These can be
understood as follows. When considering the Zeeman effect, the
density of state of electrons for the spin-up and spin-down is
different, so LLs are split into the spin-up and spin-down channels.
For a given Fermi energy $E_F$, the number of LLs below $E_F$ for
the spin-up and spin-down states can be different. If the number of
LLs below $E_F$ are even, the spin current is zero. However, if the
number of LLs below $E_F$ is odd, i.e., when $E_F$ is between the
split of the LLs, then the spin current is one half and the (Zeeman)
peak emerges in the spin current. The positions of peaks are just
the positions of LLs (see the red dotted line in fig.6(a)) since the
Zeeman split is very small (0.05t). On the other hand, in the
presence of SOI with $V_R\neq0$, the LLs for up and down spin
channels in the presence of Zeeman term are mixed together and
become two new LLs with different spin polarizations. One of LLs is
strongly spin polarized while there is no spin polarization for the
other LLs. These results are obtained by solving the
Schr$\ddot{o}$dinger equation with $V_R\neq0$. As a result, the
resonant spin current emerges when the Fermi level is just in line
with the spin-polarized LL (see the red dotted line in Fig.6(b)).
So, except for the Zeeman peaks, there are another group of peaks
which is originated from different physics in the non-zero SOI
system. From Fig.6, we can also see that the intervals of Zeeman
peaks are unchanged, while the intervals of the resonant peaks are
closer with the decreasing $V_R$ (no shown), which gives an extra
evidence that they come from the different physics.

\bigskip

\section{conclusions}

The effect of PR process $\Gamma_d$ is investigated in the 2D
four-terminal system with or without SOI. Without the SOI, the
plateaus of the Hall resistance $\rho_H$ are found to be well kept
even when the PR strength  $\Gamma_d$ is very strong (i.e. the
phase coherence length is much shorter than the size of system).
This means that the IQHE has very strong ability to resist the PR
process $\Gamma_d$. Furthermore, for the larger sample or the
stronger magnetic field, the resistive ability of the PR processes
$\Gamma_d$ is stronger. On the other hand, for the system with
SOI, the odd integer plateaus of $1/\rho_H$ are also appear. The
odd integer plateaus due to the SOI can be destroyed even for the
very weak $\Gamma_d$, but the even integer plateaus can still
survive in quite strong PR process $\Gamma_d$. Next, the SHE, i.e.
the transverse spin current, is also studied in the system with
SOI. It is found that the transverse spin current reaches the
resonant pole when the Fermi level is just consistent with the one
of the two branches of the energy eigenvalues of the system with
SOI. The PR process weakens the resonant spin Hall current and
enhances the non-resonant spin Hall current. In addition, we also
study the properties of system with the Zeeman effect and find
there are two group of peaks of the spin current originated from
different physics.

$${\bf ACKNOWLEDGMENTS}$$

We gratefully acknowledge the financial support from NSF-China under
Grant Nos. 10525418, 10734110, and 60776060, and a RGC grant from
the Government of HKSAR grant number HKU 7048/06P.


\begin{thebibliography}{00}
\bibitem{foot2}
In the real system, it is not the Fermi energy $E_F$ but the
electron density $n_s$ that is fixed while the change of the
magnetic field $B$. If the system has many impure or PR processes
and density of state is not sharp change with the energy, the
assumption of fixed the Fermi energy $E_F$ insteady of fixed the
electron density is reasonable.
\end{thebibliography}
\end{document}